\documentclass[twocolumn,amsmath,aps]{revtex4}
\usepackage{graphicx}

\newcommand{\nl}{\nonumber \\}

\newcommand{\be}{\begin{equation}}
\newcommand{\ee}{\end{equation}}
\newcommand{\bea}{\begin{eqnarray}}
\newcommand{\eea}{\end{eqnarray}}

\newcommand{\Eq}[1]{Eq.\,(\ref{#1})}

\newcommand{\la}{\langle}
\newcommand{\ra}{\rangle}
\newcommand{\dg}{\dagger}

\newcommand{\mb}{\mbox}

\begin{document}
\draft

\title{Full counting statistics of transport through two-channel
       Coulomb blockade systems}

\author{Shi-Kuan Wang}
\affiliation{State Key Laboratory for Superlattices and
Microstructures,
         Institute of Semiconductors,
      Chinese Academy of Sciences, P.O.~Box 912, Beijing 100083, China}
\author{Hujun Jiao}
\affiliation{State Key Laboratory for Superlattices and
Microstructures,
         Institute of Semiconductors,
         Chinese Academy of Sciences, P.O.~Box 912, Beijing 100083, China}

\author{Feng Li}
\affiliation{State Key Laboratory for Superlattices and
Microstructures, Institute of Semiconductors, Chinese Academy of
Sciences, P.O.~Box 912, Beijing 100083, China}

\author{Xin-Qi Li}
\email{xqli@red.semi.ac.cn} \affiliation{State Key Laboratory for
Superlattices and Microstructures,
         Institute of Semiconductors,
         Chinese Academy of Sciences, P.O.~Box 912, Beijing 100083, China}

\author{YiJing Yan}
\affiliation{Department of Chemistry, Hong Kong University of
Science and
         Technology, Kowloon, Hong Kong}

\date{\today}

\begin{abstract}
A mesoscopic Coulomb blockade system with two identical transport
channels is studied in terms of full counting statistics. It is
found that the average current cannot distinguish the quantum
constructive interference from the classical non-interference, but
the shot noise and skewness are more sensitive to the nature of
quantum mechanical interference and can fulfill that task. The
interesting super-Poisson shot noise is found and is demonstrated
as a consequence of constructive interference, which induces an
effective system with fast-and-slow transport channels. Dephasing
effects on the counting statistics are carried out to display the
continuous transition from quantum interfering to non-interfering
transports.
\end{abstract}



\maketitle

Rather than average current, the current fluctuations in
mesoscopic transport can sometimes provide deep insight into the
nature of transport mechanisms \cite{review}. A fascinating
theoretical approach, known as full counting statistics (FCS)
theory \cite{Levitov,Khlus}, can simultaneously yield all the
statistical cumulants of the number of transferred charges (i.e.,
all zero-frequency current-correlation functions). Experimentally,
the real-time counting statistics has been carried out in
transport through quantum dots \cite{single electron count},
representing a crucial achievement of being able to count
individual electron tunnel events.

For charge transport at very low transmission, the uncorrelated
transmission events are Poisson processes. However, the
Fermi-Dirac statistics together with the possible many-body
Coulomb interaction usually enhances correlation among the
transport electrons, and thereby results in sub-Poisson noise
\cite{sub-Poisson process}. It is thus of interest to examine
mechanisms that can lead to super-Poisson-noise behavior, since
the current fluctuations can be used in reverse to gain insight
into the nature of unusual transport mechanisms. The mechanisms
proposed so far for the super-Poisson noise include such as double
electron charge transfer by Andreev reflection in
normal-superconductor (NS) junction
\cite{DAR--theory,DAR--experiment}, multiple electron charge
transfer by multiple Andreev reflections in SNS system
\cite{MAR--experiment,MAR--theory1,Nazarov-circuit theory and
MAR,MAR--theory3}, dynamical channel blockade \cite{dynamical
channel blockade1,dynamical channel blockade2,dynamical channel
blockade3}, dynamical spin blockade \cite{dynamical spin
blockade}, bistability \cite{bistability mechanism}, cotunneling
\cite{cotunneling1,cotunneling2}, electron-phonon interaction in
shuttle system \cite{NEMS}, and decoherence in mesoscopic coherent
population trapping system \cite{decoherence induced}.

In this work we consider a relatively simple system, say,
electronic transport through a Coulomb blockade system with two
identical transport channels, which can be realized experimentally
by transport through two adjacent levels in a single quantum dot
(QD) \cite{Gurvitz interference}, or through two QDs in parallel
\cite{parallel double dots1,parallel double dots2}. This type of
setup itself is of particular interest, since it is an analogue of
the optical double-slit interferometer. In this context the
underlying quantum interference and phase accumulations through
QDs have been the subjects of intensive studies \cite{mesoscopic
interference review,dot in ring,Imry}. Our present study will be
placed at the level of FCS, from which a number of interesting
effects of quantum interference on current fluctuations will be
revealed.

\section*{Model Description}

The transport through a two-channel Coulomb blockade system is
governed by the Hamiltonian
\begin{subequations}
\begin{eqnarray}
& &H= H_{D}+H_{Leads}+H_{T}, \nonumber
\\
& &H_{D}=E_{1}d^{\dagger}_{1}d_{1}+E_{2}d^{\dagger}_{2}d_{2}
+Un_{1}n_{2},
\\
& &H_{Leads}=\sum_{k}(\varepsilon_{Lk} c^\dagger_{Lk}
c_{Lk}+\varepsilon_{Rk} c^\dagger_{Rk} c_{Rk}),
\\
& &H_{T}=\sum_{jk}[\Omega_{jL}d^{\dagger}_{j}c_{Lk}
+\Omega_{jR}d^{\dagger}_{j}c_{Rk}+\mb{H.c.}].
\end{eqnarray}
\end{subequations}
Here $c^\dagger_{Lk,Rk} (c_{Lk,Rk})$ and $d^{\dagger}_{j} (d_{j})$
are the electron creation (annihilation) operators, for the
electrode reservoirs and central dot states, respectively. The two
channels are characterized by states with energy levels $E_1$ and
$E_2$. Couplings of these two dot states to the electrodes are
described by $\Omega_{jL(R)}$, or $\Gamma^{j}_{L,R}=2\pi
g_{L,R}|\Omega_{jL,jR}|^{2}$, for latter use. Here $g_{L,R}$ are
the density of states (DOS) of the electron reservoirs. To
manifest maximally the quantum interference effect, we shall focus
on two \emph{identical} transmission paths. This can be
accomplished by assuming equal and energy independent coupling
strengths of the two dot states with the left and right
electrodes, i.e., $|\Omega_{1L(R)}| =
|\Omega_{2L(R)}|=\Omega_{L(R)}$, and
$\Gamma^{1}_{L(R)}=\Gamma^{2}_{L(R)}=\Gamma_{L(R)}$. To address
the quantum interference between transmissions through the two
channels, the relative phase difference is significant.
Physically, the phase difference contains the phase accumulation
of spatial motion from the electrode to dot, particularly in the
presence of magnetic vector potential (i.e., the Aharanov-Bohm
effect), as well as the phase changes associated with transmission
through quantum dots \cite{dot in ring,Imry}. Nevertheless, in
this work we would like to adopt a phenomenological way to account
for all these phase accumulations, by choosing
$\Omega_{1L}=\Omega_{2L}$, and $\eta=\Omega_{1R}/\Omega_{2R}$.
Here $\eta$ can be regarded as a relative phase parameter. Note
that the alternative gauge, say, $\Omega_{1R}=\Omega_{2R}$ and
$\eta=\Omega_{1L}/\Omega_{2L}$, does not affect the final results
\cite{Gurvitz interference}. In this paper, we also assumed
$\eta=\pm 1$, i.e., only complete constructive and destructive
interference are considered.

In the above Hamiltonian we omitted the spin indices, thus did not
explicitly write out the on-site Coulomb interaction terms, and
only left $Un_{1}n_{2}$ to describe Coulomb interaction between
electrons in the different dot states. In this work, unless
explicit specification, our study will be restricted to the strong
Coulomb blockade regime, which only allows for three available
occupation states, i.e., $|0\rangle$, $|1\rangle$, and
$|2\rangle$, corresponding to, respectively, empty dot state, and
states with one electron on either $E_{1}$ or $E_{2}$.

\section*{FCS Formulation}

Before going to the specific study of the above described system,
we would like first to reformulate the FCS formalism based on the
particle-number-resolved master equation approach
\cite{Gurvitz-n-resolved ME,Schon98,Li-PRL05,Li-PRB05}. As pointed
out in Ref.\ \onlinecite{Nazarov-PRB03}, the pioneering work
\cite{Levitov,Khlus}
 and a few other approaches developed
latter \cite{Keldysh,circuit theory,Nazarov-circuit theory and
MAR} are largely restricted to addressing the FCS of
noninteracting electrons. While Ref.\ \onlinecite{Nazarov-PRB03}
developed an elegant theory of FCS for mesoscopic systems in
strong Coulomb blockade limit, however the system's internal
\emph{quantum coherence} did not enter it since the theory was
constructed on the basis of classical stochastic processes. It is
thus advantageous to formulate an approach of being able to
account for both the internal quantum coherence and the many-body
Coulomb interaction on equal footing. Although such type of
approach has been described and applied to coherent and
interacting systems \cite{Jauho,applied in series coupled dots},
for completeness we would like here to reformulate it, hopefully
in a more transparent and unified way.

To relate with our earlier work \cite{Li-PRL05,Li-PRB05}, we
rename $H_{S}\equiv H_{D}$, $H_{B}\equiv H_{Leads}$, and reexpress
$H'\equiv H_{T}$ as
\begin{subequations}
\begin{eqnarray}
H'&=&H_{T}=\sum^{2}_{j=1}[d^{\dagger}_{j}F_{j}+\mb{H.c.}],\\
F_{j}&=&\sum_{k}\Omega_{jL}c_{Lk}+\sum_{k}\Omega_{jR}
c_{Rk}=f_{Lj}+f_{Rj}.
\end{eqnarray}
\end{subequations}
Regarding $H'$ as perturbation, the second-order cumulant
expansion leads to a formal master equation for the system's
reduced density matrix \cite{Li-PRL05, Li-PRB05}:
 \bea\label{unified master
equation} \dot{\rho}(t) &=& -i{\cal L}\rho(t) - \int^{t}_{0}d\tau
\langle{\cal L}'(t){\cal G}(t,\tau)
       {\cal L}'(\tau) {\cal G}^{\dg}(t,\tau) \rangle
       \rho(t).\notag\\
\eea Here the Liouvillian superoperators  are defined as ${\cal
L}\rho=[H_{S},\rho]$, ${\cal L}'\rho=[H',\rho]$, and ${\cal
G}(t,\tau)(...)=G(t,\tau)(...)G^{\dagger}(t,\tau)$, with
$G(t,\tau)$ the usual propagator associated with system
Hamiltonian $H_{S}$. The reduced density matrix
$\rho(t)=\mb{Tr}_{B}[\rho_{T}(t)]$, and
$\langle...\rangle=\mb{Tr}_{B}[...\rho_{B}]$ with $\rho_{B}$ the
density matrix of the electron reservoirs.

The trace in Eq.(\ref{unified master equation}) is over all the
electrode degrees of freedom. To achieve the FCS of current
fluctuations, we would like to keep track of the records of
electron numbers emitted from the source lead ($n_{1}$) and
arrived at the drain lead ($n_{2}$). We therefore classify the
Hilbert space of the reservoirs as follows: $B^{(n_1,n_2)}=
B_L^{(n_1)}\otimes B_R^{(n_2)}$. The entire Hilbert space of
electron reservoirs is thus decomposed as
$B=\bigoplus_{n_{1},n_{2}}B^{(n_1,n_2)}$.

With this classification the average over states in the entire
bath Hilbert space in \Eq{unified master equation} is replaced
with the average over states in the subspace $B^{(n_1,n_2)}$,
leading to a conditional master equation \bea\label{partial
conditional master equation} \dot{\rho}^{(n_1,n_2)}(t) &=& -i{\cal
L}\rho^{(n_1,n_2)}(t) - \int^{t}_{0}d\tau
      \mb{Tr}_{B^{(n_1,n_2)}} \nl
      & & [{\cal L}'(t){\cal G}(t,\tau)
       {\cal L}'(\tau) {\cal G}^{\dg}(t,\tau) \rho_T(t)] .
\eea Here,
$\rho^{(n_1,n_2)}(t)=\mb{Tr}_{B^{(n_1,n_2)}}[\rho_T(t)]$, is the
reduced density matrix of the central system conditioned by the
electron numbers emitted from the source lead ($n_{1}$) and
arrived at the drain lead ($n_{2}$) until time $t$.

To proceed, following Ref.\ \onlinecite{Li-PRB05}, two physical
considerations are further implemented: (i) Instead of the
conventional Born approximation for the entire density matrix
$\rho_{T}(t)\simeq \rho(t)\bigotimes\rho_{B}$, we propose the
ansatz
$\rho_T(t)\simeq\sum_{n_1,n_2}\rho^{(n_1,n_2)}(t)\otimes\rho_B^{(n_1,n_2)}$,
where $\rho_B^{(n_1,n_2)}(t)$ is the density operator of the
electron reservoirs associated with $n_{1}$-electrons emitted from
the source and $n_{2}$-electrons entered the drain. The
orthogonality between reservoirs states in different subspaces
leads to the term selection from the entire density operator
$\rho_{T}$. (ii) Due to the closed nature of the transport
circuit, the extra electrons entered the drain will flow back into
the source via the external circuit. Also the rapid relaxation
processes in the reservoirs will quickly bring the reservoirs to
local thermal equilibrium characterized by the chemical
potentials. As a consequence, after the state selection procedure,
the electron reservoirs density matrices $\rho^{(n_{1},n_{2})}$
should be replaced by $\rho^{(0)}_{B}$.

Further use of the Markov-Redfield approximation leads
Eq.(\ref{partial conditional master equation}) to an explicit
form:
\begin{subequations}\label{n-ME}
\bea &&\dot{\rho}^{(n_1,n_2)}=  -i {\cal L}\rho^{(n_1,n_2)}
      - \frac{1}{2}{\cal R}_1\rho^{(n_1,n_2)},\\
&&{\cal R}_1\rho^{(n_1,n_2)}  = \sum_{j}
    [d_{j}^{\dg}A_{j}^{(-)}\rho^{(n_1,n_2)}
   +\rho^{(n_1,n_2)}A_{j}^{(+)}d_{j}^{\dg}    \nl
 & & ~~~~~~~~        - A_{Lj}^{(-)}\rho^{(n_1-1,n_2)}d_{j}^{\dg}
            - d_{j}^{\dg}\rho^{(n_1+1,n_2)}A_{Lj}^{(+)}        \nl
 & & ~~~ - A_{Rj}^{(-)}\rho^{(n_1,n_2-1)}d_{j}^{\dg}
   - d_{j}^{\dg}\rho^{(n_1,n_2+1)}A_{Rj}^{(+)}]+{\rm H.c.}~.     \nl
\eea
\end{subequations}
Here $A_{\alpha j}^{(+)}=\sum_{i} \tilde{C}_{\alpha ij}^{(+)}(+
{\cal L})d_{i}$, $A_{\alpha j}^{(-)}=\sum_{i} \tilde{C}_{\alpha
ji}^{(-)}(- {\cal L})d_{i}$,  and
$A_{j}^{(\pm)}=\sum_{\alpha=L,R}A_{\alpha j}^{(\pm)}$. The
spectral functions are defined in terms of the Fourier transform
of the reservoir correlation functions, i.e.,
$\tilde{C}^{(\pm)}_{\alpha ij}(\pm
\mathcal{L})=\int^{\infty}_{-\infty}dt C^{(\pm)}_{\alpha
ij}(t)e^{\pm i\mathcal{L}t}$. The reservoir correlators read $\la
f_{\alpha i}^{\dg}(t)f_{\alpha j}(\tau)\ra = C_{\alpha
ij}^{(+)}(t-\tau)$, and $\la f_{\alpha i}(t)f^{\dg}_{\alpha
j}(\tau)\ra = C_{\alpha ij}^{(-)}(t-\tau)$. Here $\la \cdots \ra$
stands for $\mb{Tr}_B [\cdots\rho_B^{(0)}]$, with the usual
meaning of thermal average. Obviously, $\la
F_{i}^{\dg}(t)F_{j}(\tau)\ra =
C_{ij}^{(+)}(t-\tau)=\sum_{\alpha=L,R}C_{\alpha
ij}^{(+)}(t-\tau)$, and $\la F_{i}(t)F^{\dg}_{j}(\tau)\ra =
C_{ij}^{(-)}(t-\tau) =\sum_{\alpha=L,R}C_{\alpha
ij}^{(-)}(t-\tau)$. For the sake of brevity, the explicit
expressions of the reservoir correlation functions, the
corresponding spectral functions, and $A_{\alpha j}^{(\pm)}$ are
ignored here, and are presented alternatively in Appendix A.

At this stage, it is worth making a few remarks as follows: (i)
The above particle-number-resolved master equation is applicable
to finite temperatures, which is an extension of Gurvitz's
approach \cite{Gurvitz-n-resolved ME}. (ii) The second-order
cumulant expansion of $H'$ restricts the applicability to the
regime of sequential tunneling. However, generalization to higher
order expansion of $H'$ \cite{high-order
expansion--Yan1,high-order expansion--Yan2}
 and self-consistent corrections\cite{Li--self
consistence} are possible. The corresponding FCS version is an
interesting subject for future work. (iii) The above
$(n_1,n_2)$-resolved master equation generalizes the result in
Ref.\ \onlinecite{Li-PRB05}, for electron counting from at one
junction to at two junctions. Further generalization to
multi-terminal setup is straightforward, following precisely the
same treatment. (iv) The connection of the
particle-number-resolved density matrix with the distribution
function of FCS is obvious, i.e.,
$P(n_1,n_2,t)=\mb{Tr}[\rho(t)^{(n_1,n_2)}]$, where the trace is
over the central system states. From this distribution function,
all orders of cumulants of transmission electrons can be
calculated.

In practice, instead of obtaining the distribution function from
the solution of the particle-number-resolved master equation, a
more efficient method is the cumulant generating function (CGF)
technique. In the following study, we only consider single
counting statistics. That is, we only keep $n_2$, after making
summation over $n_1$. Multiple counting statistics in
multi-terminal setup follows the same technique.

Mathematically, the CGF is introduced as
\begin{equation}
e^{-F(\chi)}=\sum_n P(n,t)e^{in\chi} .
\end{equation}
Here $\chi$ corresponds to the so-called counting field. Based on
the CGF, the $k_{\rm th}$ cumulant reads
$C_k=-(-i\partial_{\chi})^k F(\chi)|_{\chi=0}$. For instance, the
first two cumulants give rise to the mean value of the transmitted
electron numbers $C_1=\bar{n}$, and the variance
$C_2=\overline{n^2}-\bar{n}^2$; the third cumulant (skewness),
$C_3=\overline{(n-\bar{n})^3}$, characterizes the asymmetry of the
distribution. Here, $\overline{(\cdots)}=\sum_n (\cdots)P(n,t)$.
Moreover, the cumulants are straightforwardly related to the
transport characteristics, e.g., the average current by $I =
eC_{1}/t$, and the zero-frequency shot noise by $S =
2e^{2}C_{2}/t$. The Fano factor is defined as $F=C_{2}/C_{1}$,
which represents the amplitude of current fluctuations, with $F>1$
indicating a super-Poisson fluctuation, and $ F<1$ a sub-Poisson
process.

The generating function can be calculated as follows. Define
$S(\chi,t)=\sum_n \rho^{(n)}(t)e^{in\chi}$. Obviously,
$e^{-F(\chi)}={\rm Tr}[S(\chi,t)]$. Let us reexpress the
particle-number-resolved master equation formally as \bea
\dot{\rho}^{(n)}=A\rho^{(n)}+C\rho^{(n+1)}+D\rho^{(n-1)}, \eea
then $S(\chi,t)$ satisfies \bea
\dot{S}=AS+e^{-i\chi}CS+e^{i\chi}DS \equiv {\cal L}_{\chi} S .
\eea The formal solution reads $S(\chi,t)=e^{{\cal
L}_{\chi}t}S(\chi,0)$. In the low frequency limit, the counting
time is much longer than the time of tunneling through the system.
One can prove \cite{Nazarov-PRB03, applied in series coupled
dots,decoherence induced,Jauho}, that
$F(\chi)=-\lambda_{1}(\chi)t$, where $\lambda_{1}(\chi)$ is the
eigenvalue of $\mathcal{L}_{\chi}$, and satisfies the condition
$\lambda_{1}(\chi)\mid_{\chi\rightarrow 0}\rightarrow 0$.

\section*{FCS Analysis}

We now turn to the specific system under study. In the strong
Coulomb blockade regime and at zero temperature, the matrix
element form of the particle-number-resolved master equation reads
\begin{subequations}\label{n-EQ-1}
\begin{eqnarray}
\dot{\rho}^{(n_{2})}_{00}&=&-2\Gamma_{L}\rho^{(n_{2})}_{00}
+\Gamma_{R}(\rho^{(n_{2}-1)}_{11}+\rho^{(n_{2}-1)}_{22})\nonumber\\
& &+\eta\Gamma_{R}(\rho^{(n_{2}-1)}_{12}+\rho^{(n_{2}-1)}_{21}),
\\
\dot{\rho}^{(n_{2})}_{11}&=&\Gamma_{L}\rho^{(n_{2})}_{00}
-\Gamma_{R}\rho^{(n_{2})}_{11}-\eta\frac{\Gamma_{R}}{2}(\rho^{(n_{2})}_{12}+\rho^{(n_{2})}_{21}),
\\
\dot{\rho}^{(n_{2})}_{22}&=&\Gamma_{L}\rho^{(n_{2})}_{00}
-\Gamma_{R}\rho^{(n_{2})}_{22}-\eta\frac{\Gamma_{R}}{2}(\rho^{(n_{2})}_{12}+\rho^{(n_{2})}_{21}),
\\
\dot{\rho}^{(n_{2})}_{12}&=&i\delta\epsilon\rho^{(n_{2})}_{12}
+\Gamma_{L}\rho^{(n_{2})}_{00}-\Gamma_{R}\rho^{(n_{2})}_{12}\nonumber\\&
&-\eta\frac{\Gamma_{R}}{2}(\rho^{(n_{2})}_{11}+\rho^{(n_{2})}_{22}).
\end{eqnarray}
\end{subequations}
Here we have summed $n_1$ and remained only $n_2$, indicating the
mere study of FCS of the electrons entered the drain reservoir.
For clarity, we have denoted the level spacing by
$\delta\epsilon=E2-E1$, and choose the reference of zero energy
such that $E2=\delta\epsilon/2$, and $E1=-\delta\epsilon/2$. In
the derivation of Eqs. (\ref{n-EQ-1}), we assumed that $E_{1,2}$
are inside the window of bias voltage,
i.e.,$\mu_{L}>E_{1,2}>\mu_{R}$, and $U$ is infinite. In Ref.\
\onlinecite{Gurvitz interference} the same master equation was
derived by a wavefunction approach.

Performing a discrete Fourier transformation
$\sum_{n_{2}}e^{in_{2}\chi}$ to Eqs.(\ref{n-EQ-1}), we obtain
\begin{equation}\label{k-ME-2}
\mathcal{L}_{\chi} =
\begin{pmatrix}
-2\Gamma_{L} &\Gamma_{R}e^{i\chi}&\Gamma_{R}e^{i\chi}
&\eta\Gamma_{R}e^{i\chi}&\eta\Gamma_{R}e^{i\chi}\\ \Gamma_{L}&
-\Gamma_{R}&0&-\eta\frac{\Gamma_{R}}{2}&-\eta\frac{\Gamma_{R}}{2}\\
\Gamma_{L}&0&-\Gamma_{R}
&-\eta\frac{\Gamma_{R}}{2}&-\eta\frac{\Gamma_{R}}{2}\\
\Gamma_{L}&-\eta\frac{\Gamma_{R}}{2}&-\eta\frac{\Gamma_{R}}{2}&i\delta\epsilon-\Gamma_{R}&0\\
\Gamma_{L}&-\eta\frac{\Gamma_{R}}{2}&-\eta\frac{\Gamma_{R}}{2}&0&-i\delta\epsilon-\Gamma_{R}
\end{pmatrix}
\end{equation}
According to the definition of the cumulants we can express
$\lambda_{1}(\chi)$ as
\begin{equation}
\lambda_{1}(\chi)=\frac{1}{t}\sum^{\infty}_{k=1}C_k\frac{(i\chi)^{k}}{k!}
.
\end{equation}
Then insert the above expansion into
$|\lambda_{1}(\chi)I-\mathcal{L}_{\chi}|=0$, and expand this
determinant in series of $(i\chi)^{k}$. Since the value of $i\chi$
is arbitrary we can obtain $C_k$ by setting the coefficients of
$(i\chi)^{k}$ equal to zero and solving them sequentially
\cite{decoherence induced}. Analytic expressions of the first two
cumulants are accordingly obtained as
\begin{subequations}
\begin{eqnarray}
C_{1}&=&\frac{2\Gamma_{L}\Gamma_{R}\delta\epsilon^{2}}{2\Gamma_{L}
(\delta\epsilon^{2}-(\eta-1)\Gamma^2_{R})+\Gamma_{R}\delta\epsilon^{2}},\\
C_{2}&=&\frac{\delta\epsilon^{4}\Gamma^{2}_{R}+4\Gamma^{2}_{L}
[\delta\epsilon^4+2\delta\epsilon^2
\eta\Gamma^{2}_{R}+(\eta-1)^2\Gamma^{4}_{R}]}{[2\Gamma_{L}
(\delta\epsilon^{2}-(\eta-1)\Gamma^2_{R})+\Gamma_{R}\delta\epsilon^{2}]^3}\nonumber\\&
&\times 2\Gamma_{L}\Gamma_{R}\delta\epsilon^{2}.
\end{eqnarray}
\end{subequations}
while the higher order cumulants can be instead carried out
numerically, to avoid their lengthy expressions.

\begin{figure}
\begin{center}
\includegraphics*[width=8cm,height=7cm,keepaspectratio]{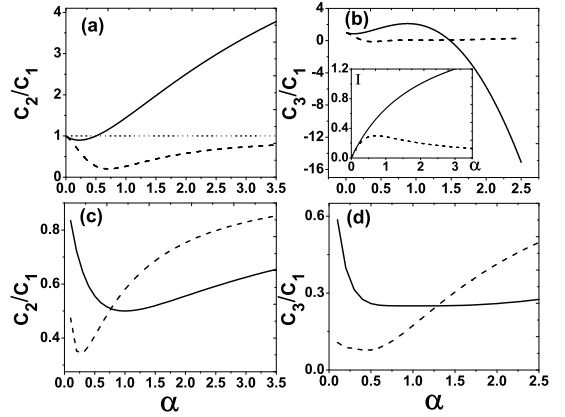}
\caption{\label{mainresults} First three cumulants of the
zero-frequency current fluctuations. The solid and dashed curves
display the results for constructive ($\eta=1$) and destructive
($\eta=-1$) interferences, respectively. The variable
$\alpha=\Gamma_{R}/\Gamma_{L}$ characterizes the asymmetry of
dot-state couplings to the left and right electrodes. (a) and (b)
show the Fano factor and skewness for strong Coulomb blockade
system, while (c) and (d) for system allowing double occupancy,
i.e., $U=0.0$, as a comparison. Inset of (b) plots the average
current $I = eC_{1}/t$, with the convention $e=1$. In the
calculation $\Gamma_{L}=1.0\delta\epsilon$ was assumed.}
\end{center}
\end{figure}

In Fig.\ 1 the first three cumulants of transport current are
displayed. It is of interest to note that in the Coulomb-blockade
regime a super-Poisson noise is developed by the {\it constructive
interference} between the two paths (i.e. $\eta=+1$). This is
clearly shown by the solid curve in Fig.\ 1(a). With the increase
of the coupling asymmetry (i.e. $\alpha=\Gamma_R/\Gamma_L$), the
super-Poisson feature will be more evident. In contrast, for
destructive interference ($\eta=-1$), the current fluctuation is
sub-Poissonian, as plotted by the dashed curve in Fig.\ 1(a).

Another intriguing finding is that the super- and sub-Poisson
characteristics are associated with different behaviors of the
skewness $C_{3}/C_{1}$, as shown in Fig.\ 1(b). For destructive
interference, the skewness is approximately zero, meanwhile for
constructive interference, transition of the skewness from (small)
positive to (large) negative values takes place, by increasing the
coupling asymmetry ($\alpha$). As is well known for photon
counting statistics in quantum optics, the skewness (both its
magnitude and sign) provides further information for the counting
statistics, beyond the second order cumulant. As a comparison, the
results of non-interacting system are plotted in Fig.1 (c) and
(d), where neither the super-Poisson noise nor the negative
skewness is found.

To understand better the above super-Poisson behavior, below we
present an analysis in terms of {\it fast-and-slow} transport
channels. Let us introduce an alternative representation for the
double dot states, with the corresponding electronic operators
$f_{1}=(\Omega_{L}d_{1}
+\Omega_{R}d_{2})/\sqrt{\Omega_{L}^{2}+\Omega_{R}^{2}}$, and
$f_{2}=(\Omega_{L}d_{2}-\Omega_{R}d_{1}
)/\sqrt{\Omega_{L}^{2}+\Omega_{R}^{2}}$, as well as the state
energies $E_{1/2}= \mp \frac{\delta\epsilon}{2}$. In such
representation the entire Hamiltonian is reexpressed as
\begin{equation}
\begin{split}
H&=\tilde{E}_{1}f^{\dagger}_{1}f_{1}
+\tilde{E}_{2}f^{\dagger}_{2}f_{2}
+\gamma(f^{\dagger}_{1}f_{2}+f^{\dagger}_{2}f_{1})
\\
&+\sum_{k}[\tilde{\Omega}_{1L}f^{\dagger}_{1}a_{Lk}
+\tilde{\Omega}_{1R}f^{\dagger}_{1}a_{Rk}+\mb{H.c.}]
\\
&+\sum_{k}[\tilde{\Omega}_{2L}f^{\dagger}_{2}a_{Lk}
+\tilde{\Omega}_{2R}f^{\dagger}_{2}a_{Rk}+\mb{H.c.}] ~.
\end{split}
\end{equation}
Note that the strong Coulomb blockade, rather than being
explicitly described in this Hamiltonian, is reflected
alternatively by the single occupation of the two dot states. For
the sake of brevity, explicit expressions for the coupling between
the new dot states, and their couplings to the electrodes are
presented in Appendix B.

\begin{figure}
\begin{center}
\includegraphics*[width=8cm,height=7cm,keepaspectratio]{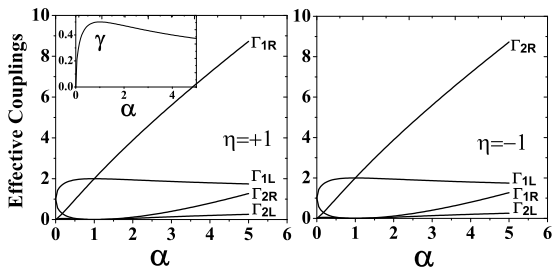}
\caption{\label{fastslwochannel} Effective couplings of the dot
states to the electrodes in the transformed state representation.
It is found that effective fast-and-slow transport channels are
evaluated for the constructive interference ($\eta=1$), which is
contrasted remarkably with the destructive interference
($\eta=-1$). Inset: effective coupling between the transformed dot
states. Parameters $\alpha=\Gamma_{R}/\Gamma_{L}$, and
$\Gamma_{L}=\delta\epsilon=1.0$.   }
\end{center}
\end{figure}

In the new states representation, the formation of the
fast-and-slow transport channels is demonstrated in Fig.\ 2: (i)
for constructive interference ($\eta=+1$), increasing
$\alpha=\Gamma_R/\Gamma_L$ can lead to $\Gamma_{1L}>\Gamma_{2L}$,
and $\Gamma_{1R}>>\Gamma_{2R}$; (ii) for destructive interference
($\eta=-1$), however, increasing $\alpha=\Gamma_R/\Gamma_L$ leads
to $\Gamma_{1L}>\Gamma_{2L}$, but $\Gamma_{2R}>>\Gamma_{1R}$. 
Obviously, in the former case, an effective fast-and-slow
transport channels are evaluated, but {\it can not} in the second
case. Such fast-and-slow transport channels will lead to bunching
behavior, and result in the super-Poisson noise \cite{dynamical
channel blockade1,dynamical channel blockade2,dynamical channel
blockade3,dynamical spin blockade}.

\section*{Decoherence Effect}

To illustrate the decoherence effect, we consider dephasing
between the two interfering paths, which is described by including
the matrix element form of the following Lindblad-type terms
\begin{equation}
L_{\phi i}\rho^{(n_{2})}L^{\dagger}_{\phi i}
-\frac{1}{2}L^{\dagger}_{\phi i}L_{\phi i}\rho^{(n_{2})}
-\frac{1}{2}\rho^{(n_{2})} L^{\dagger}_{\phi i}L_{\phi i}
\nonumber
\end{equation}
into Eq.(9), where the jump operators $L_{\phi
1}=\sqrt{\Gamma_{d}}|1\rangle\langle 1|$, and $L_{\phi
2}=\sqrt{\Gamma_{d}}|2\rangle\langle 2|$. The effects of
decoherence on the first three cumulants are shown in Fig.\ 3,
where the solid and dashed curves correspond to the constructive
($\eta=1$) and destructive ($\eta=-1$) interferences,
respectively. Interestingly, for the constructively interfering
transport ($\eta=1$), dephasing does not influence the transport
current, see the solid line in Fig.\ 3(a). This is in remarkable
contrast with the result of double-slit optical interference,
where the constructively interfering intensity is four times of
the intensity of the individual path (slit), while the
non-interfering intensity is simply two times of the single-slit
intensity. This essential difference is originated from the
multiple forward-and-backward scattering between the dot states
and the electrodes in the case of electron transport. However,
also for $\eta=1$, the second and third cumulants (i.e. $C_2$ and
$C_3$) sensitively depend on the dephasing strength. In
particular, dephasing would cause a transition from super-Poisson
to sub-Poisson processes, meanwhile the skewness ($C_3$) changes
from negative value to zero. The present analysis clearly shows
that the super-Poisson current fluctuation is a consequence of the
constructive interference. For destructively interfering transport
($\eta=-1$), the almost vanished transport current will be
restored by dephasing, whereas the shot noise and skewness
approximately do not change with dephasing.

\begin{figure}
\begin{center}
\includegraphics*[width=8cm,height=7cm,keepaspectratio]{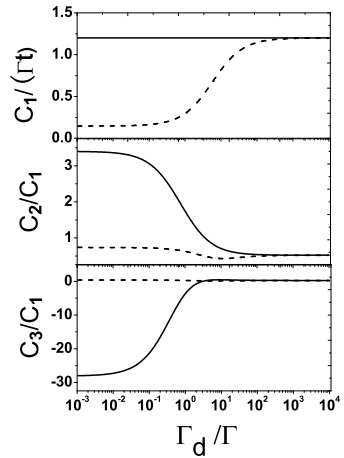}
\caption{\label{decoherence} Dephasing effects on the first three
cumulants, which show a continuous transition from quantum
interfering to classical non-interfering transports. The solid and
dashed curves display results for the constructive ($\eta=1$) and
destructive ($\eta=-1$) interferences, respectively. Parameters
$\alpha=\Gamma_{R}/\Gamma_{L}=3.0$, and
$\Gamma_{L}=\delta\epsilon=\Gamma$.   }
\end{center}
\end{figure}

\section*{Conclusion}

To summarize, we have presented a FCS study for transport through
a mesoscopic Coulomb blockade system with two identical transport
channels. The FCS analysis showed that the shot noise and skewness
are more sensitive than the average current to the underlying
quantum interference. In particular, the average current cannot
distinguish the quantum constructive interference from the
classical non-interference, while the shot noise and skewness can
fulfill that task. In the regime of quantum constructive
interference, the interesting super-Poisson shot noise was found,
and was understood in terms of an effective fast-and-slow channel
picture via state-representation transformation. Dephasing effect
has also been carried out to display the continuous transition of
the first three cumulants from quantum interfering to classical
non-interfering transports. Experiments within current technology
are capable of examining the predictions of this work.

Methodologically, the present work extended the orthodox FCS
approach in strong Coulomb-blockade systems \cite{Nazarov-PRB03}.
The present approach can not only account for the many-body
Coulomb interaction, but also handle the internal quantum
coherence. This advantage allows for a wider range of
applications.


\vspace{3ex} {\it Acknowledgments.} This work was supported by the
National Natural Science Foundation of China under grant No.\
60425412 and No.\ 90503013, and the Research Grants Council of the
Hong Kong Government.

\appendix
\section{}

In this appendix we present the explicit expressions of the
spectral functions in Eq.\ (5) in the main text. For
noninteracting electrodes and under the wide-band approximation,
the reservoir correlation functions simply read
\begin{subequations}
\begin{eqnarray}
C^{(\pm)}_{L ij}(t-\tau)&=&g_{L} \Omega_{L}^{2}\int
d\varepsilon_{L k}e^{\pm i\varepsilon_{L
k}(t-\tau)}n^{(\pm)}_{L}(\varepsilon_{L k}),\notag\\\\
C^{(\pm)}_{R ii}(t-\tau)&=&g_{R} \Omega_{R}^{2}\int
d\varepsilon_{R k}e^{\pm i\varepsilon_{R
k}(t-\tau)}n^{(\pm)}_{R}(\varepsilon_{R k}),\notag\\\\
C^{(\pm)}_{R 12}(t-\tau) &=&\eta g_{R} \Omega_{R}^{2}\int
d\varepsilon_{R k}e^{\pm i\varepsilon_{R
k}(t-\tau)}n^{(\pm)}_{R}(\varepsilon_{R k})\notag\\&=&C^{(\pm)}_{R
21}(t-\tau)
\end{eqnarray}
\end{subequations}
Here, the indices $i$ and $j$ denote the dot-states $|1\ra$ and
$|2\ra$. Since the two dot-states are almost degenerate in energy
and are equally coupled to the electrodes, in the above we have
assumed that $\Omega_{i\alpha }\Omega^*_{j\alpha }
=\Omega^*_{i\alpha }\Omega_{j\alpha }=\Omega_{\alpha}^2$. Note
that an exception is $\Omega^*_{1R }\Omega_{2R }=\Omega^*_{2R
}\Omega_{1R }=\eta \Omega_{R}^2$, where $\eta=\pm 1$ is the
relative phase factor. This is because we have attributed the
phase difference to the coupling amplitudes with the right
electrodes. The electron and hole occupation functions are
introduced, respectively, as $n^{(+)}_{\alpha}(\varepsilon_{\alpha
k})=n_{\alpha}(\varepsilon_{\alpha k})$, and
$n^{(-)}_{\alpha}(\varepsilon_{\alpha
k})=1-n_{\alpha}(\varepsilon_{\alpha k})$, with
$n_{\alpha}(\varepsilon_{\alpha k})$ the Fermi function. Fourier
transformation of Eq.\ (A1) gives the spectral functions:
\begin{subequations}
\begin{eqnarray}
\tilde{C}^{(\pm)}_{L ij}(\omega)&=&\Gamma_{L}
n^{(\pm)}_{L}(\mp\omega),
\\
\tilde{C}^{(\pm)}_{R ii}(\omega)&=&\Gamma_{R}
n^{(\pm)}_{R}(\mp\omega),
\\
\tilde{C}^{(\pm)}_{R 12}(\omega)&=&\tilde{C}^{(\pm)}_{R
21}(\omega)=\eta\Gamma_{R} n^{(\pm)}_{R}(\mp\omega),
\end{eqnarray}
\end{subequations}
where $\Gamma_{\alpha}=2\pi g_{\alpha} \Omega_{\alpha}^{2}$.

Furthermore, using  $\mathcal{L}d_{j}=-(E_{j}+Un_{\bar{j}})d_{j}$,
the operators $A_{\alpha j}^{(\pm)}$ in Eq.\ (5), which is defined
by $A_{\alpha j}^{(+)}=\sum_{i} \tilde{C}_{\alpha ij}^{(+)}(+
{\cal L})d_{i}$, $A_{\alpha j}^{(-)}=\sum_{i} \tilde{C}_{\alpha
ji}^{(-)}(- {\cal L})d_{i}$, are accordingly obtained as
\begin{subequations}
\begin{eqnarray}
A^{(+)}_{\alpha j}&=&\sum_{i}\tilde{C}^{(+)}_{\alpha
ij}[-(E_{i}+Un_{\bar{i}})]d_{i},
\\
A^{(-)}_{\alpha j}&=&\sum_{i}\tilde{C}^{(-)}_{\alpha
ji}[+(E_{i}+Un_{\bar{i}})]d_{i}.
\end{eqnarray}
\end{subequations}
Here the index $\bar{i}$ simply means differing from $i$, i.e.,
$\bar{i}=2$ if $i=1$, and vice versa.


\section{}

Via the transformation of state-representation as described in the
main text, the energy levels of the transformed dot-states and
their effective coupling strength read
\begin{equation}
\begin{split}
\tilde{E}_{1}&=\frac{\delta\epsilon}{2}\frac{\Omega^{2}_{R}
-\Omega^{2}_{L}}{\Omega^{2}_{L}+\Omega^{2}_{R}},\\
\tilde{E}_{2}&=\frac{\delta\epsilon}{2}\frac{\Omega^{2}_{L}
-\Omega^{2}_{R}}{\Omega^{2}_{L}+\Omega^{2}_{R}},
\\
\gamma&=\delta\epsilon\frac{\Omega_{L}\Omega_{R}}{\Omega^{2}_{L}+\Omega^{2}_{R}}.
\end{split}
\end{equation}
Simple algebra also gives rise to the effective coupling strengths
of the dot-states with the electrodes:
\begin{equation}
\begin{split}
\tilde{\Omega}_{1L}&=\frac{\Omega_{1L}}{\sqrt{\Omega^{2}_{L}
+\Omega^{2}_{R}}}(\Omega_{L}+\Omega_{R}),
\\
\tilde{\Omega}_{1R}&=\frac{\Omega_{2R}}{\sqrt{\Omega^{2}_{L}
+\Omega^{2}_{R}}}(\eta\Omega_{L}+\Omega_{R}),
\\
\tilde{\Omega}_{2L}&=\frac{\Omega_{1L}}{\sqrt{\Omega^{2}_{L}
+\Omega^{2}_{R}}}(\Omega_{L}-\Omega_{R}),
\\
\tilde{\Omega}_{2R}&=\frac{\Omega_{2R}}{\sqrt{\Omega^{2}_{L}
+\Omega^{2}_{R}}}(\Omega_{L}-\eta\Omega_{R}).
\end{split}
\end{equation}
More transparently, for $\eta=1$, the corresponding tunneling
rates read
\begin{equation}
\begin{split}
\Gamma_{1L}&=\frac{\Gamma_{L}(\Gamma_{L}+\Gamma_{R}
+2\sqrt{\Gamma_{L}\Gamma_{R}})}{\Gamma_{L}+\Gamma_{R}},\\
\Gamma_{1R}&=\frac{\Gamma_{R}(\Gamma_{L}+\Gamma_{R}
+2\sqrt{\Gamma_{L}\Gamma_{R}})}{\Gamma_{L}+\Gamma_{R}},\\
\Gamma_{2L}&=\frac{\Gamma_{L}(\Gamma_{L}+\Gamma_{R}
-2\sqrt{\Gamma_{L}\Gamma_{R}})}{\Gamma_{L}+\Gamma_{R}},\\
\Gamma_{2R}&=\frac{\Gamma_{R}(\Gamma_{L}+\Gamma_{R}
-2\sqrt{\Gamma_{L}\Gamma_{R}})}{\Gamma_{L}+\Gamma_{R}},
\end{split}
\end{equation}
and for $\eta=-1$, they are
\begin{equation}
\begin{split}
\Gamma_{1L}&=\frac{\Gamma_{L}(\Gamma_{L}+\Gamma_{R}
+2\sqrt{\Gamma_{L}\Gamma_{R}})}{\Gamma_{L}+\Gamma_{R}},\\
\Gamma_{1R}&=\frac{\Gamma_{R}(\Gamma_{L}+\Gamma_{R}
-2\sqrt{\Gamma_{L}\Gamma_{R}})}{\Gamma_{L}+\Gamma_{R}},\\
\Gamma_{2L}&=\frac{\Gamma_{L}(\Gamma_{L}+\Gamma_{R}
-2\sqrt{\Gamma_{L}\Gamma_{R}})}{\Gamma_{L}+\Gamma_{R}},\\
\Gamma_{2R}&=\frac{\Gamma_{R}(\Gamma_{L}+\Gamma_{R}
+2\sqrt{\Gamma_{L}\Gamma_{R}})}{\Gamma_{L}+\Gamma_{R}}.
\end{split}
\end{equation}

\end{document}